\documentclass[]{raa}            % referee version: for submission
\usepackage{graphicx,times}
\usepackage{natbib}
\usepackage{amsmath}
\usepackage{float}
\usepackage{subfigure}
\usepackage{indentfirst} %首行缩进
\newcommand{\unit}[1]{\ensuremath{\,\mathrm{#1}}} 
\providecommand{\bm}[1]{\mbox{\boldmath$#1$\unboldmath}}

\begin{document}
   \title{Dependence of the Length of Solar Filament Threads on the Magnetic
	  Configuration}
 \volnopage{ {\bf 201X} Vol.\ {\bf X} No. {\bf XX}, 000--000}
   \setcounter{page}{1}
   \author{Y. H. Zhou \inst{1,2} \and
           P. F. Chen \inst{1,2} \and
           Q. M. Zhang \inst{3}  \and
	   C. Fang     \inst{1,2}
}
 \institute{School of Astronomy and Space Science, Nanjing University,
             Nanjing 210093, China; {\it yuhaozhou1991@gmail.com;
	     chenpf@nju.edu.cn}\\
            \and
	Key Laboratory of Modern Astronomy \& Astrophysics,
        Nanjing University, Nanjing 210093, China \\
	    \and
	Purple Mountain Observatory, Chinese Academy of Sciences, Nanjing
	210008, China\\
\vs \no
   {\small Received [year] [month] [day]; accepted [year] [month] [day] }
}

\abstract{
High-resolution H$\alpha$ observations indicate that filaments consist of an
assembly of thin threads. In quiescent filaments, the threads are generally
short, whereas in active region filaments, the threads are generally long. In
order to explain these observational features, we performed one-dimensional
radiative hydrodynamic simulations of filament formation along a dipped 
magnetic flux tube in the framework of the chromospheric evaporation-coronal
condensation model. The geometry of a dipped magnetic flux tube is characterized by three parameters, i.e., the depth ($D$), the half-width ($w$), and the altitude ($h$) of the magnetic dip. The parameter survey in the numerical simulations shows that allowing the filament thread to grow in 5 days, the maximum length ($L_{th}$) of the filament thread increases linearly with $w$, and decreases linearly with $D$ and $h$. The dependence is fitted into a linear function $L_{th}=0.84w-0.88D-2.78h+17.31$ Mm. Such a relation can qualitatively explain why quiescent filaments have shorter threads and active region filaments have longer threads.
\keywords{Sun:filaments, prominences - Methods: numerical - Hydrodynamics}
}

\authorrunning{Zhou, Chen, Zhang, \& Fang }            
\titlerunning{Length of Solar Filament Threads}  
\maketitle

\section{Introduction} 

Filaments, which are also called solar prominences when they appear above the
solar limb, are cold and dense plasma concentrations in the corona 
\citep{tand95}. They often appear as a narrow spine above the magnetic
polarity inversion line \citep{zirk89, mart98, berg08, ning09}. Because of the
large density and hence the large gravity, filament threads are often
considered to be supported by the magnetic tension force of the dip-shaped
magnetic loops either in a normal-polarity configuration \citep{ks57} or in an
inverse-polarity configuration \citep{kr74}. Whereas the corresponding global
magnetic configuration in the latter case is typically a flux rope as
derived from coronal magnetic field extrapolations \citep[e.g.,][]{su12}, the
magnetic configuration in the former case might be just a sheared arcade
system \citep{chen12}, although theoretical models of flux ropes with the
normal-polarity configuration have also proposed \citep[e.g.,][]{zhan05}.

Despite that it has been studied for decades, the formation of filaments is
still a hot research topic in solar physics and has been discussed in a variety
of works since filaments and the host magnetic structure are thought to be the
progenitor of coronal mass ejections \citep{chen11}. High-resolution
observations have shown that the filament spine is composed of a collection of
separate threads \citep{engv04, lin05}. These threads are believed to be the
building blocks of filaments. So, to understand the formation of a filament,
the formation of a filament thread should be explained in the first place. A filament thread is generally thought to be located along a flux tube. The
magnetic field of quiescent or intermediate filaments is generally weak, e.g.,
several Gauss, while that of active region filaments are much stronger, e.g.,
tens of Gauss \citep{aula03}. Therefore, for active region filaments, the
formation of their threads can be simplified into a one-dimensional (1D)
radiative hydrodynamic problem. Even for intermediate filaments, the 1D 
numerical simulation made by \citet{zhan12} can still reproduce the oscillation
behaviors revealed by satellite observations. Therefore, 1D hydrodynamic model
is still a good approximation even for intermediate filaments. More 
importantly, the easy-going 1D radiative hydrodynamic simulations can provide straightforward insight into the physics related to the formation and 
oscillation processes in solar filaments. For example, with 1D radiative hydrodynamic simulations, \citet{ant99} proposed the chromospheric evaporation plus coronal condensation model, where extra heating localized in the chromosphere drives chromospheric evaporation into the corona, and the dense hot plasma near the magnetic dip cools down to form a filament thread due to thermal instability.

In the chromospheric evaporation plus coronal condensation model, it was found
that with the intermediately asymmetric heating at the two footpoints of a flux tube, a filament thread forms and then drains down to the chromosphere repetitively. If the chromospheric heating at the two footpoints is relatively symmetric or the magnetic dip is deep enough, the filament thread would be held near the magnetic dip, growing continually \citep{Karp01}. Recently, \citet{xia11} 
performed state-of-the-art radiative hydrodynamic simulations of the filament
thread formation process. It was found that once the coronal condensation 
happens, no further chromospheric heating is needed for the filament thread
to grow. The siphon flow induced by the pressure imbalance between the
chromosphere and the condensation would supply more and more plasma into the
coronal condensation, leading to the continual growth of the filament thread.
As the plasma accumulates, the gas pressure within the filament would 
increase, which hinders the siphon flow from the chromosphere. If so, an
interesting question is raised, that is, for a given magnetic flux tube, what
is the maximum length that a filament thread can grow. In this paper, we aim 
to investigate how long the filament thread would finally be and how the
maximum length is related to the geometry of the magnetic flux tube. Our
numerical method is introduced in Section \ref{sect2}, the results are 
presented in Section \ref{sect3}, which are discussed in Section \ref{sect4}.

\section{Numerical Method}
\label{sect2}

Similar to what we have done before \citep{xia11, zhan12, zhang12, zhan13}, we deal with 1D radiative hydrodynamic equations, where the magnetic field is not taken into account explicitly under the assumption that the plasma dynamics does not affect the magnetic field. Note that the shape of the magnetic flux tube is represented by the distribution of the field-aligned component of the gravity, $g_\parallel$. The hydrodynamic equations shown  below are numerically solved by the state-of-the-art MPI-Adaptive Mesh Refinement-Versatile Advection Code \citep[MPI-AMRVAC,][]{Kepp03, Kepp12}:

\begin{equation}
\frac{{\partial \rho }}{{\partial t}} + \frac{\partial }{{\partial s}}(\rho v) = 0,
\label{eq1}
\end{equation}
\begin{equation}
\frac{\partial }{{\partial t}}(\rho v) + \frac{\partial }{{\partial s}}(\rho {v^2} + p) = \rho {g_\parallel}(s),
\label{eq2}
\end{equation}
\begin{equation}
\frac{{\partial E}}{{\partial t}} + \frac{\partial }{{\partial s}}(Ev + pv) = \rho {g_\parallel}v  - {n_e}{n_H}\Lambda (T) + \frac{\partial }{{\partial s}}(\kappa \frac{{\partial T}}{{\partial s}})+ H(s),
\label{eq3}
\end{equation}

\noindent
where $\rho$ is the mass density, $s$ is the distance along the magnetic loop
starting from the left footpoint, $v$ is the velocity, $p$ is the gas pressure,
$T$ is the temperature, and ${g_\parallel}(s)$ is the field-aligned component
of the gravity at the distance $s$ along the magnetic loop, which is
determined by the geometry of the magnetic loop. Besides, $E = \rho {v^2}/2 +
p/(\gamma-1)$ is the total energy density, where $\gamma  = 5/3$ is the
adiabatic index. The second term on the right-hand side of Eq. (\ref{eq3}) is
the optically thin radiative cooling, where ${n_e}$ is the number density of
electrons, ${n_H}$ the number density of hydrogen, and $\Lambda (T)$ the
radiative loss coefficient. The third term on the right-hand side of Eq. 
(\ref{eq3}) is the heat conduction, where $\kappa  = {10^{ - 6}}{T^{5/2}}
\unit{ergs}\unit{cm}^{-1}\unit{s}^{-1}\unit{K}^{-1}$ is the Spitzer-type heat
conductivity. The last term on the right-hand side of Eq. (\ref{eq3}), $H(s)$,
is the volumetric heating rate. It includes the steady background heating 
$H_0$ and the localized chromospheric heating, whose expressions will be
discussed at the end of this section. We assume a fully ionized plasma model
and take $\rho  = 1.4m_pn_H$ and $p = 2.3n_Hk_BT$ considering the helium
abundance ($n_e/n_H=0.1$), where $m_p$ is the proton mass and ${k_B}$ is the
Boltzmann constant. To calculate the radiative energy loss coefficient 
$\Lambda (T)$, a second-order polynomial interpolation is taken to compile a 
high resolution table based on the radiative loss calculations using an 
accurate atomic collisional rate and a recommended set of quiet-region element 
abundances over a wide temperature range \citep{colg08}.

\begin{figure}[h]
\centering
\includegraphics[width=0.6\textwidth]{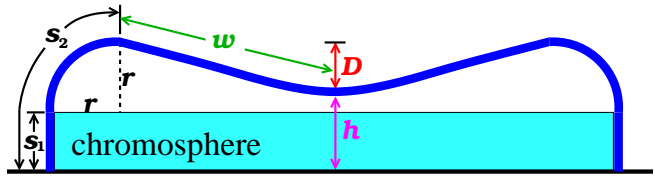}
\caption{Geometry of the magnetic loop used for the 1D radiative hydrodynamic
   simulations of the filament formation, where the blue thick line is the
   magnetic loop which is anchored to the solar chromosphere. Note that the
   horizontal and the vertical sizes are not to scale.}\label{fig1}
\end{figure}	

As mentioned before, it is widely believed that a filament is hosted at the
dip of a magnetic loop, supported by its magnetic tension force. So, we adopt
a symmetric loop including a magnetic dip, as shown in
Fig. \ref{fig1}. The loop, whose total length is $L$, is composed of two
vertical legs with a length of ${s_1}$ for each, two quarter-circular
shoulders with a radius $r$, and a quasi-sinusoidal-shaped dip with a 
half-length of $w$. The dip has a depth of $D$ below the apex of the loop. Then, we define $s_2$ as ${s_2} = s_1 + \pi r/2$, the altitude of the dip 
($h$) as $h = s_1 + r - D$, and the total length of the dip $2w$ as $2w = L - 2{s_2}$. The corresponding $g_\parallel$ in Eqs. 
(\ref{eq2}--\ref{eq3}) is then expressed as follows:

\begin{equation}
	g_{\parallel}=
	\begin{cases}
		- {g_ \odot }, & s \le {s_1}; \\
		- {g_ \odot }\cos \left( {\frac{\pi }{2}\frac{{s - {s_1}}}{{{s_2} - {s_1}}}} \right) & {s_1} < s \le {s_2};\\
		{g_ \odot }\frac{{\pi D}}{{2(L/2 - {s_2})}}\sin \left( {\pi \frac{{s - {s_2}}}{{L/2 - {s_2}}}} \right) & {s_2} < s \le {L/2};
	\end{cases}
\end{equation}
where ${g_ \odot } = 2.7 \times {10^2}\unit{m}\unit{s}^{-2}$ is the gravitational acceleration at the solar surface.

Our simulations start from a thermal and force-balanced equilibrium state. Initially, the background heating $H_0$ is included in order to balance the
thermal conduction and radiative cooling. The plasma in the loop is static.
Then, each simulation is divided into two steps: 1) Filament formation:
Localized chromospheric heating $H_1$ is introduced symmetrically near the
footpoints of the magnetic loop so that the chromospheric material is
evaporated into the corona and later condensates at a certain stage due to
thermal instability, and then a filament thread forms and grows at the center
of the magnetic loop; 2) Relaxation: The localized heating $H_1$ is ramped 
down to zero linearly after a decay timescale of 1000 s, and the chromospheric
evaporation ceases. Owing to the disappearance of the evaporation flow, the
compressed filament thread relaxes and expands. After that, the filament thread grows in length slowly due to the self-induced siphon flow as found by \citet{xia11}. The formation and relaxation time will be described in Sect. \ref{siphon}.

In step 1, the volumetric heating rate $H(s)$ in Eq. (\ref{eq3}) is composed
of two parts: the steady background heating $H_0(s)$ and the localized chromospheric heating $H_1(s)$. Their expressions are:

\begin{equation}
	{H_0}(s)=
	\begin{cases}
		{E_0}\exp ( - s/{H_m}) & s < L/2;\\
		{E_0}\exp [ - (L - s)/{H_m}] & L/2 \le s < L;\\
	\end{cases}
\end{equation}

\begin{equation}
	{H_1}(s)=
	\begin{cases}
		{E_1} & s \le s_{tr}\\
		{E_1}\exp [ - (s - {s_{tr}})/\lambda ] & s_{tr} < s \le L/2;\\
		{E_1}\exp [ - (L - {s_{tr}} - s)/\lambda ] & L/2 < s \le L-s_{tr};\\
		{E_1} & s>L-s_{tr}.
	\end{cases}
\end{equation}

The background heating term $H_0(s)$ is adopted to maintain the 1 MK corona with the amplitude ${E_0} = 3 \times {10^{ - 4}}\unit{ergs}\unit{cm^{-3}}\unit{s^{-1}}$, and the scale-height ${H_m}$ is defined as ${H_m}=L/2$. The localized heating term $H_1$ is adopted to generate chromospheric evaporation into the corona with the amplitude ${E_1} = 3 \times {10^{ - 2}}\unit{ergs}\unit{cm^{-3}}\unit{s^{-1}}$. The height of the transition region, $s_{tr}$, is set to be $6\unit{Mm}$, and the scale-height $\lambda$ is $10\unit{Mm}$. 

\section{Numerical Results}\label{sect3}

\subsection{Natural growth via siphon flows}
\label{siphon}

\begin{figure}[h]
	\centering
	\subfigure{%\subfigure[Step1]
	{\includegraphics[width=0.4\textwidth]{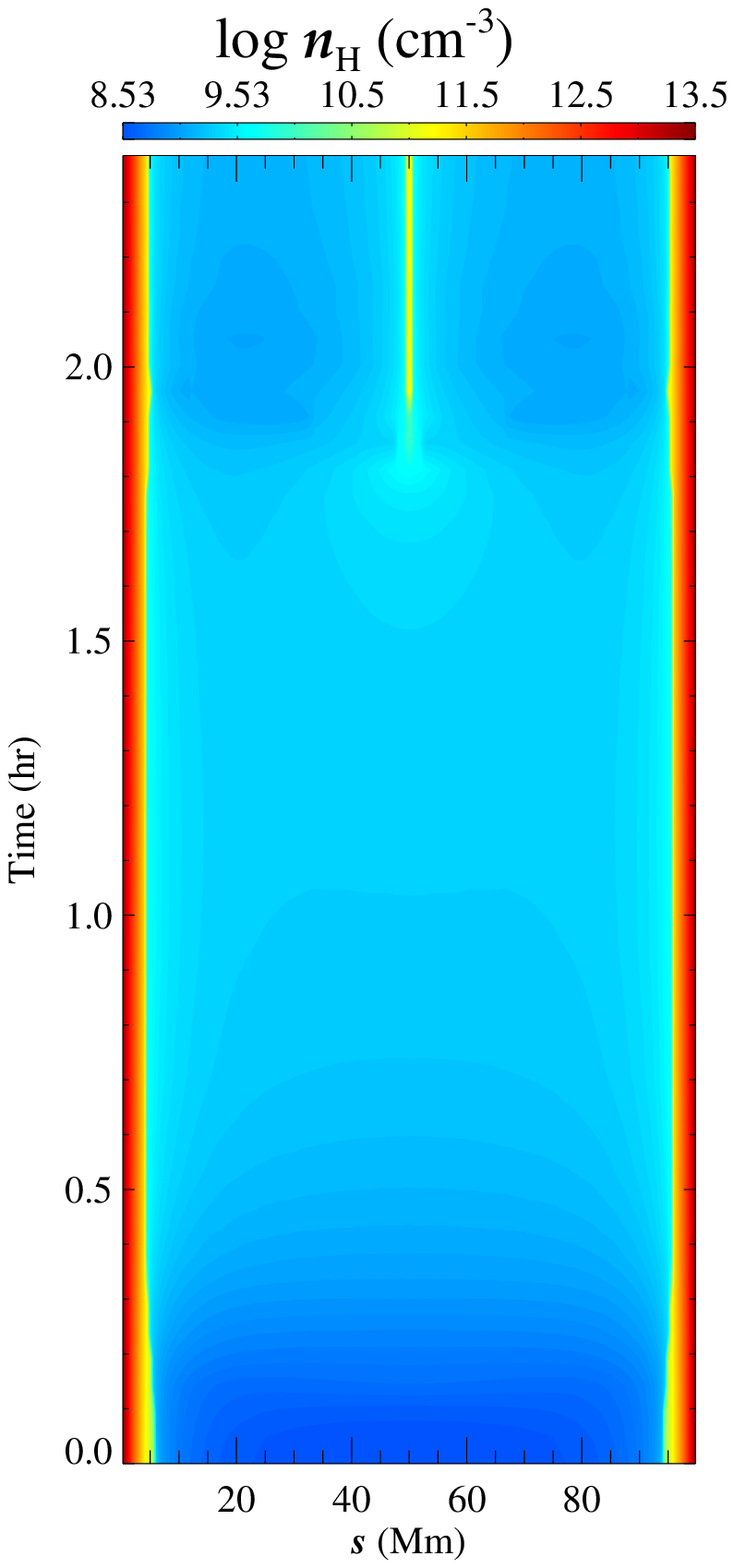}}}
	\subfigure{%\subfigure[Step2]
	{\includegraphics[width=0.4\textwidth]{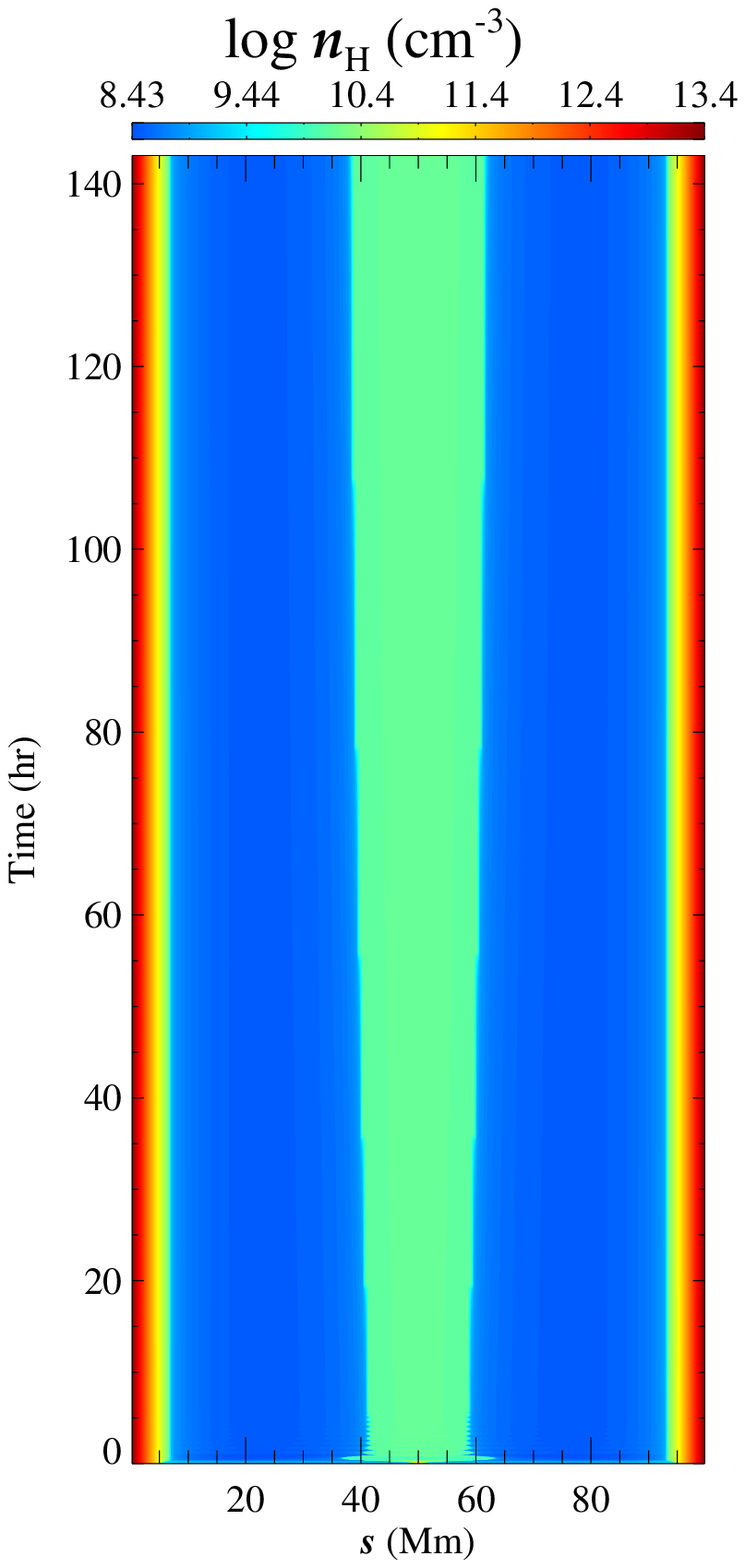}}}	
	\caption{Time evolution of the density distribution along the magnetic loop during the two stages: (a) the filament formation stage; (b) the self-growth stage.}\label{fig2}
\end{figure}

According to \citet{xia11}, once coronal condensation happens, the filament
thread can grow via siphon flows even without further localized chromospheric
heating. In order to check how long the filament thread can grow, we perform
a simulation of filament formation with the similar parameters used in 
\citet{xia11}, i.e., $s_1$=5 Mm, $r=5$ Mm, $2w=74.3$ Mm, $D=1$ Mm, and $L=100$ Mm. The evolution of the density distribution along the magnetic loop in step 1 is shown in the left panel of Fig. \ref{fig2}. It is found that as localized heating $H_1$ is introduced near the two footpoints, the coronal part of the loop becomes hotter and denser. After 2 hrs, thermal instability occurs, and a segment of filament thread is formed as indicated by the high density near the loop center. As the chromospheric evaporation goes on, the filament grows with a rate of 0.2 km s$^{-1}$. Such a simulation in step 1 continues until the filament thread grows for 1 hr. The length of the filament thread is $\sim$3 Mm. Then in step 2, we ramp down the localized chromospheric heating $H_1$ to zero in 1000 seconds. The corresponding evolution of the density distribution along the magnetic loop in step 2 is depicted in the right panel of Fig. \ref{fig2}. It is seen that after the localized heating is switched off, the filament thread suddenly expands and then shrinks. Such an oscillation decays rapidly, and the filament thread soon reaches a quasi-equilibrium state, with a length of about 16 Mm. After that, the length of the filament thread increases slowly.

\begin{figure}[h]
	\centering{
	\includegraphics[width=0.7\textwidth]{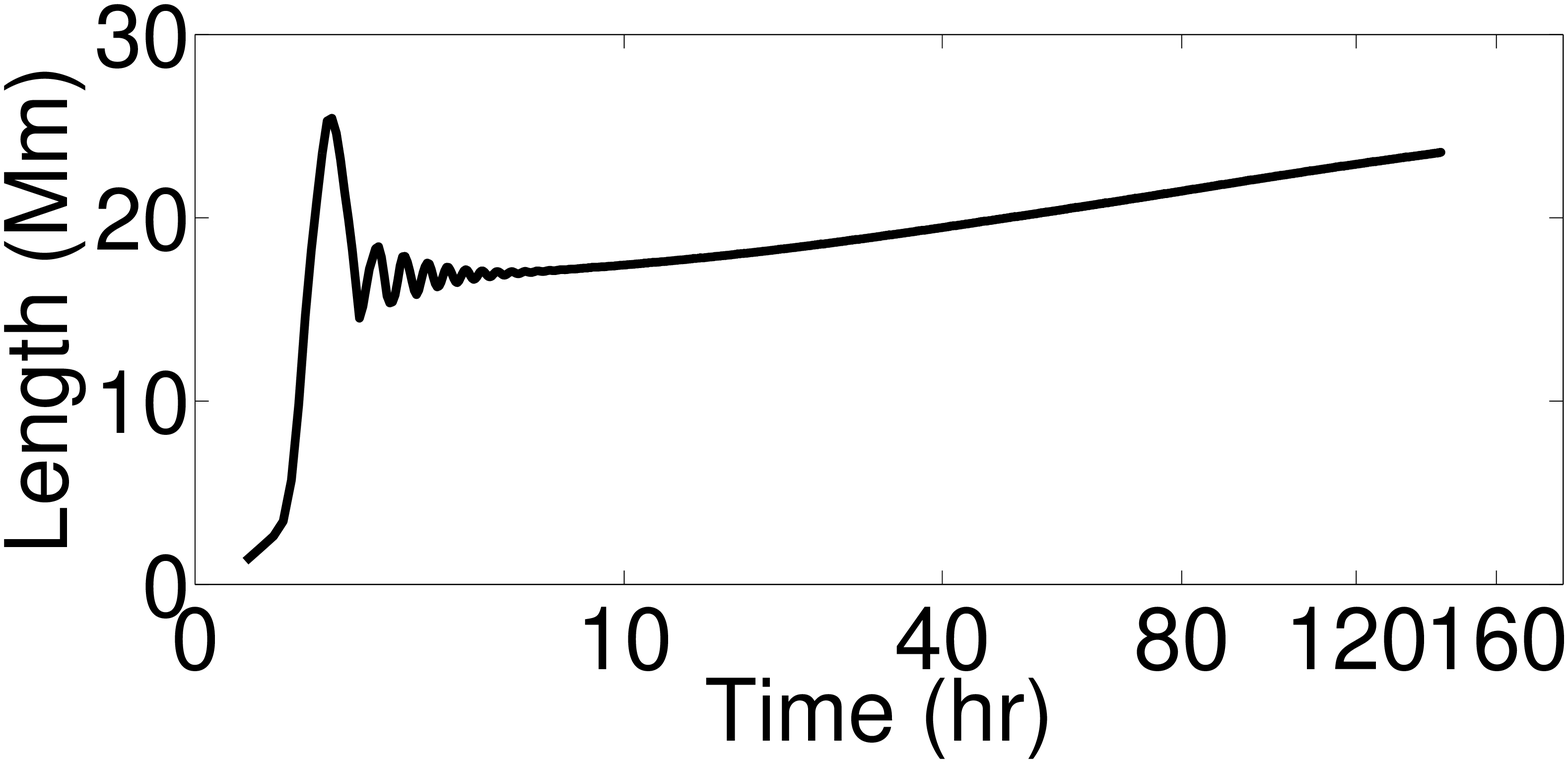}}
	\caption{Time evolution of the length of the filament thread in the relaxation stage.}\label{fig3}
\end{figure}

Fig. \ref{fig3} shows the growth of the length of the filament thread with time in step 2.
It is found that as the localized heating and the resulting evaporation are switched off, the length of the filament thread first increases rapidly to 24 Mm, and then oscillates around 17 Mm. The oscillation decays rapidly, and then the length of the thread begins to increase slowly. It is seen that the growth rate decreases with time with an initial rate of 70 km h$^{-1}$. After about 120 hr, the filament thread is 22.8 Mm long, and its growth rate becomes about 0.7 Mm per day,
which means that the filament thread expands $\sim$1\arcsec\ per day. Such a
rate is nearly imperceptible in observations in visual inspection. So we can consider that the filament thread length saturates at 22.8 Mm. 

\subsection{Parameter survey}

We apply the method mentioned in Sect. \ref{siphon} to investigate how the maximum length of the filament thread ($L_{th}$) changes with the geometric parameters of the magnetic flux tube, i.e., the
depth ($D$), the half-width ($w$), and the altitude ($h$) of the magnetic dip.

\begin{figure}[h]
	\centering
	\includegraphics[width=0.5\textwidth]{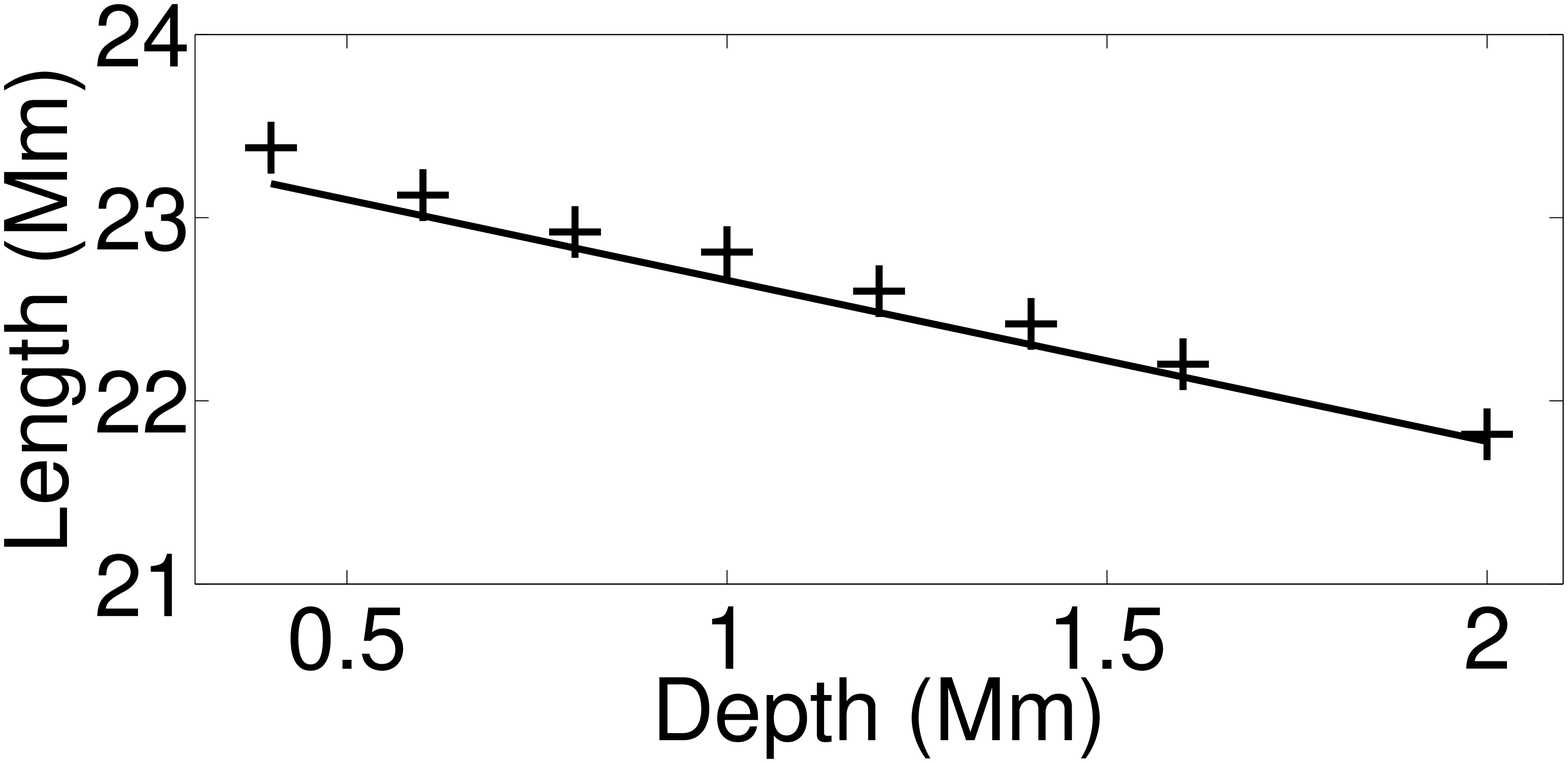}
	\caption{Dependence of the length of the filament thread on the depth of the magnetic dip. The plus signs are from the numerical simulations, while the straight line represents the function which is fit for the whole parameter survey.}\label{fig4}
\end{figure}

By fixing $w=$ 37.2 Mm and $h$= 9 Mm, eight cases with $D$ ranging from 0.4 Mm to
2.0 Mm are simulated. The variation of the filament thread length ($L_{th}$)
along with the depth of the magnetic dip ($D$) is shown in Fig. \ref{fig4}.
It is revealed that $L_{th}$ decreases nearly linearly with increasing $D$. The scatter plot in Fig. \ref{fig4} can be fit with a linear function $L_{th} = 23.6 - 0.91D$ (Mm). It is noted as $D$ increases from 0.4 Mm to 2 Mm, the length of the filament thread decreases $\sim$10\% only. 

\begin{figure}%%%%[h]
	\centering
	\includegraphics[width=0.5\textwidth]{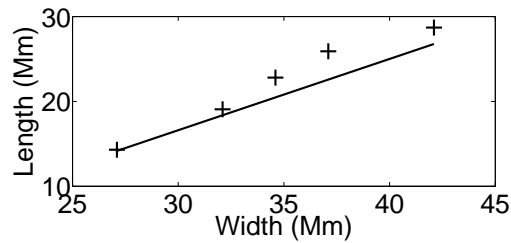}
	\caption{Dependence of the length of the filament thread on the width of the magnetic dip. The plus signs are from the numerical simulations, while the straight line represents the function which is fit for the whole parameter survey.}\label{fig5}
\end{figure}

By fixing $D=$ 1 Mm and $h$= 9 Mm, five cases with $w$ ranging from 27.1 Mm to
42.1 Mm are simulated. The variation of the filament thread length ($L_{th}$)
along with the width of the magnetic dip ($D$) is shown in Fig. \ref{fig5}.
It is revealed that $L_{th}$ increases linearly with increasing $w$, i.e., the
filament thread length is longer for a wider magnetic dip. Their relation can
be fit with a function, $L_{th} = 0.82w - 9.20$ (Mm). Such a straight line intersects with the $x$-axis at $w=11.2$ Mm, implying that under the conditions with $D=$ 1 Mm and $h$= 9 Mm there exists such a threshold of the width of the magnetic dip, $w=11.2$ Mm, below which no filament can be formed. This is consistent with the criterion of thermal instability \citep{Park53} qualitatively. 

\begin{figure}%%%[H]
	\centering
	\includegraphics[width=0.5\textwidth]{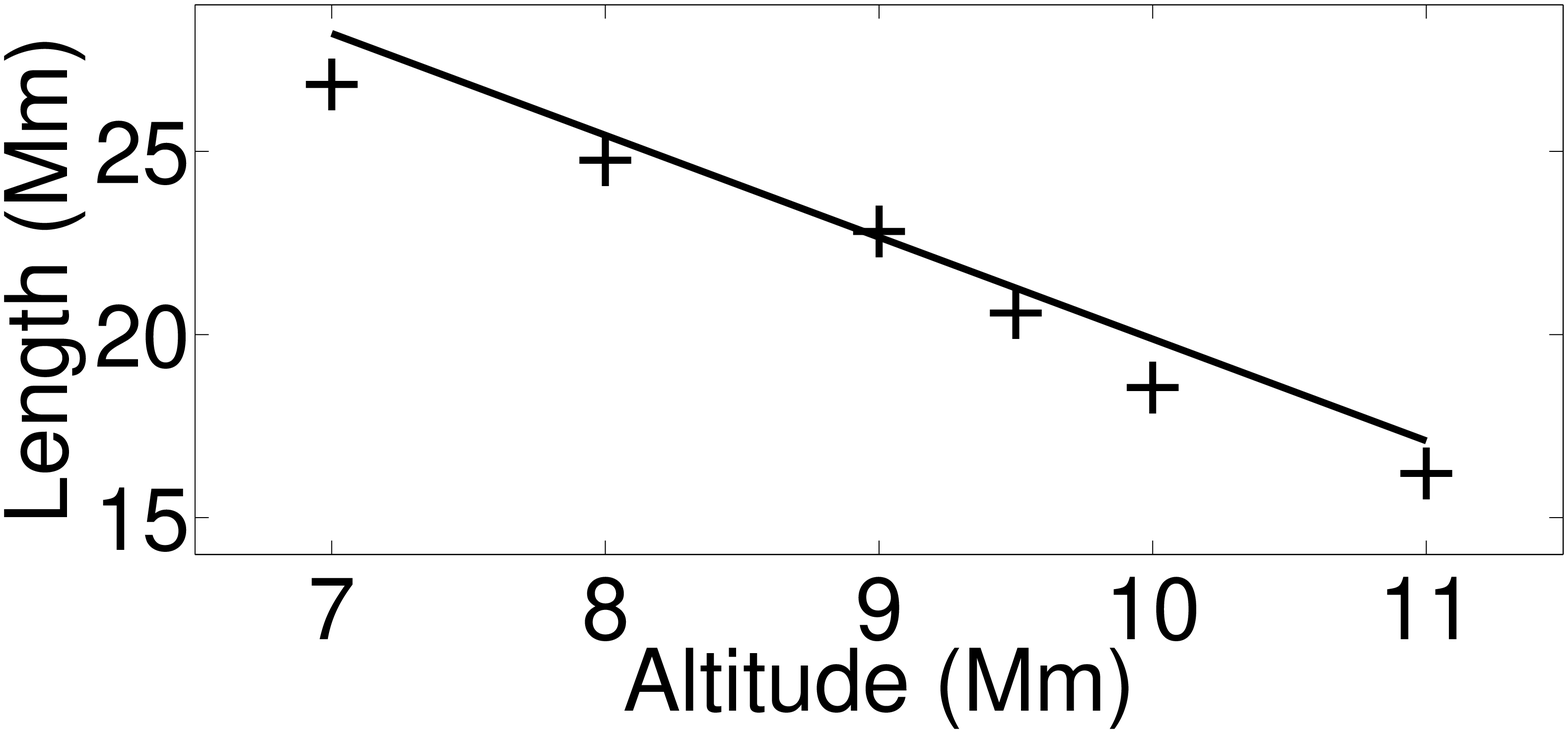}
	\caption{Dependence of the length of the filament thread on the altitude of the magnetic dip. The plus signs are from the numerical simulations, while the straight line represents the function which is fit for the whole parameter survey.}\label{fig6}
\end{figure}

By fixing $D=$ 1 Mm and $w$= 37.2 Mm, five cases with $h$ ranging from 7 Mm to
12 Mm are simulated. The variation of the filament thread length ($L_{th}$)
along with the altitude of the magnetic dip ($h$) is shown in Fig. \ref{fig6}.
It is revealed that $L_{th}$ decreases with increasing $h$, i.e., the filament
thread length is shorter for a higher magnetic dip. Their relation can be fit
with a function, $L_{th} = 43.56 - 2.48h$ (Mm). Such a straight line intersects with the $x$-axis at $h=17.6$ Mm, implying that under the conditions with $D=$ 1 Mm and $w$= 37.2 Mm there exists such an upper limit of the altitude of the magnetic dip, $h=17.6$ Mm, above which no filament can be formed.

\section{Discussions}\label{sect4}

More and more observational evidence tends to support the
evaporation-condensation model for the filament formation \citep[e.g.,][]
{liu12, berg08}. Correspondingly, several groups have performed radiative
hydrodynamic simulations, mainly in 1D, to confirm the model \citep[e.g.,][]{ant99, Karp01, luna12, xia12}. Recently, \citet{xia12} extended the
simulations from 1D to 2D. All their results revealed that as localized heating
is introduced in the chromosphere, the local plasma is evaporated into the 
corona, a cool filament is formed when the criterion of the thermal instability is satisfied. The filament thread grows as the evaporation continues. Even when the chromospheric evaporation is switched off, the filament thread can still grow via siphon flows \citep{xia11}. However, as the siphon flows accumulate in the filament, its gas pressure will increase, which in turn hinders the siphon flows. Therefore, a crucial question is when the filament thread will stop growing. To answer this question, for the first time we performed 1D hydrodynamic simulations, which proceed until the length of the filament thread saturates in a prescribed flux tube with a magnetic dip. The parameter survey indicates that the maximum length ($L_{th}$) of a filament thread is strongly related to the half-width ($w$), the depth ($D$) and the altitude ($h$) of the magnetic dip. Based on the numerical results,
we derived a function for the dependence on each parameter. Since all the functions are linear, we describe the dependence of the length of filament threads with a multi-variable linear function, i.e., $L_{th}=\alpha_1 w +\alpha_2 D + \alpha_3 h+\alpha_4$. The least square fit to all the data points in Figs. \ref{fig4}--\ref{fig6} yields 
$$L_{th}=(0.84\pm 0.07)w-(0.88\pm 0.27)D-(2.78\pm 0.17)h+(17.31\pm 3.07),$$
\noindent
where all the terms are in units of Mm. Such a fitted linear function is plotted in Figs. \ref{fig4}--\ref{fig6} as straight lines.

Observations indicate that a filament is composed of a collection of threads
\citep{lin05}. In quiescent filaments, the threads are generally short, 
whereas in active region filaments, the threads are relatively long 
\citep{mack10}. These observations can qualitatively explained by the simulation
results presented in this paper: Active regions are born to be non-potential, implying the field lines are initially strongly sheared. Besides, active regions often experience fast rotations \citep{yan08}, which would further drag the magnetic field lines to become elongated,
and the dips become shallow, i.e., the magnetic dips of active region field
lines correspond to a large width ($w$) and a small depth ($D$). According to
our formula, the corresponding filament threads would be long. On the
other hand, quiescent filaments are formed in decayed active regions 
\citep{vanb89}. Therefore, the corresponding magnetic fields might have a 
shorter $w$ and larger $D$. As a result, the filament threads in quiescent
filaments are shorter than those in active region filaments. Observations also
indicate that quiescent filaments are higher located in the corona, e.g., 
$10^4$--$10^5$ km, whereas active region filaments are often below $10^4$ km in
altitude \citep{mack10}. Our result of the dependence of the filament
thread length on the altitude of the magnetic dip implies that the longer
threads of active region filaments are partly due to the low altitude of the
magnetic dip.

It is noted that the simulations in this paper are aimed to study the length for a filament thread to grow in a dipped magnetic flux tube in 5 days. In real
observations, the lifetime of some filaments might be less than 5 days before they erupt when a certain instability happens \citep{chen08, shen11}. Besides, there is continual mass drainage from the filament to the chromosphere \citep{xia11, liu12}, a filament thread in observations may have not yet reached its upper limit. Even though, as discussed above, the dependence of the filament thread length on the altitude and the width of the magnetic dip can still account for the typical characteristics of both quiescent and active region filaments qualitatively. In order to compare the simulation results with H$\alpha$ observations more quantitatively, we plan to (1) do the extrapolation of the coronal force-free field based on the magnetogram observations, then (2)
perform a series of 1D hydrodynamic simulations along many field lines
retrieved from the extrapolated, and then (3) compare the simulated thread
assembly with the H$\alpha$ observations made by our ONSET telescope \citep{fang13} in the partial-disk mode. This will be another step forward beyond \citet{aula99} and \citet{guna13}. It will also be interesting to check whether our scaling law for the length of the filament thread is applicable to the feet (or barbs) of the filaments as more and more observations of barbs have been made \citep[e.g.,][]{li13}.

\normalem
\begin{acknowledgements}
The authors thank the referee for useful comments. The research is supported by
the Chinese foundations NSFC (11025314, 10878002, 10933003, and 11173062) and
2011CB811402.
\end{acknowledgements}

\label{lastpage}


\begin{thebibliography}{99}
\small \setlength{\itemindent}{-3mm} \setlength{\itemsep}{-0.5mm}
\setlength{\baselineskip}{4.5mm}

\bibitem[Antiochos et al.(1999)]{ant99} Antiochos, S.~K., MacNeice, P.~J., 
	Spicer, D.~S., \& Klimchuk, J.~A.\ 1999, \apj, 512, 985 
\bibitem[Aulanier \& D{\'e}moulin(2003)]{aula03} Aulanier, G., \& D{\'e}
	moulin, P.\ 2003, \aap, 402, 769 
\bibitem[Aulanier et al.(1999)]{aula99} Aulanier, G., D{\'e}moulin, P., Mein, 
	N., et al.\ 1999, \aap, 342, 867 
\bibitem[Berger et al.(2008)]{berg08} Berger, T.~E., Shine, R.~A., Slater, 
	G.~L., et al.\ 2008, \apjl, 676, L89
\bibitem[Chen(2011)]{chen11} Chen, P.~F.\ 2011, Living Reviews in Solar
	Physics, 8, 1
\bibitem[Chen(2012)]{chen12} Chen, P.~F.\ 2012, Hinode-3: The 3rd Hinode
	Science Meeting, 454, 265 
\bibitem[Chen et al.(2008)]{chen08} Chen, P.~F., Innes, D.~E., \& Solanki,
	S.~K.\ 2008, \aap, 484, 487 
\bibitem[Colgan et al.(2008)]{colg08} Colgan, J., Abdallah, J., Jr., Sherrill,
	M.~E., et al.\ 2008, \apj, 689, 585 
\bibitem[Engvold(2004)]{engv04} Engvold, O.\ 2004, Multi-Wavelength
	Investigations of Solar Activity, 223, 187
\bibitem[Fang et al.(2013)]{fang13} Fang, C., Chen, P.~F., Li, Z., et al.\
	2013, \raa, 13, 1509 
\bibitem[Gun{\'a}r et al.(2013)]{guna13} Gun{\'a}r, S., Mackay, D.~H., 
	Anzer,	U., \& Heinzel, P.\ 2013, \aap, 551, A3
\bibitem[Karpen et al.(2001)]{Karp01} Karpen, J.~T., Antiochos, S.~K.,
	Hohensee, M. et al.\ 2001, \apjl, 553, L85 
\bibitem[Keppens et al.(2003)]{Kepp03} Keppens, R., Nool, M., Toth, G., \&
	Goedbloed, J. P. \ 2003, Comput. Phys., 153, 317
\bibitem[Keppens et al (2012)]{Kepp12} Keppens, R., Meliani, Z., van Marle,
	A. J. et al. \ 2012, J. Comput. Phys., 231, 718
\bibitem[Kippenhahn \& Schl{\"u}ter(1957)]{ks57} Kippenhahn, R., \& 
	Schl{\"u}ter, A.\ 1957, \zap, 43, 36 
\bibitem[Kuperus \& Raadu(1974)]{kr74} Kuperus, M., \& Raadu, M.~A.\ 1974,
	\aap, 31, 189
\bibitem[Li \& Zhang(2013)]{li13} Li, L., \& Zhang, J.\ 2013, \solphys, 282,
	147 
\bibitem[Lin et al.(2005)]{lin05} Lin, Y., Engvold, O., Rouppe van der Voort, 
	L., Wiik, J.~E., \& Berger, T.~E.\ 2005, \solphys, 226, 239
\bibitem[Liu et al.(2012)]{liu12} Liu, W., Berger, T.~E., \& Low, B.~C.\ 2012,
	\apjl, 745, L21
\bibitem[Luna \& Karpen(2012)]{luna12} Luna, M., \& Karpen, J.\ 2012, \apjl, 
	750, L1 
\bibitem[Mackay et al.(2010)]{mack10} Mackay, D.~H., Karpen, J.~T., Ballester, 
	J.~L., Schmieder, B., \& Aulanier, G.\ 2010, \ssr, 151, 333 
\bibitem[Martin(1998)]{mart98} Martin, S.~F.\ 1998, \solphys, 182, 107 
\bibitem[Ning et al.(2009)]{ning09} Ning, Z., Cao, W., \& Goode, P.~R.\ 
	2009, \apj, 707, 1124
\bibitem[Parker(1953)]{Park53}{Parker}, E.~N. 1953, \apj, 117, 431
\bibitem[Shen et al.(2011)]{shen11} Shen, Y.-D., Liu, Y., \& Liu, R.\ 2011,
	\raa, 11, 594
\bibitem[Su \& van Ballegooijen(2012)]{su12} Su, Y., \& van Ballegooijen, A.\
	2012, \apj, 757, 168
\bibitem[Tandberg-Hanssen(1995)]{tand95} Tandberg-Hanssen, E.\ 1995,
	Astrophysics and Space Science Library, 199
\bibitem[van Ballegooijen \& Martens(1989)]{vanb89} van Ballegooijen, A.~A.,
	\& Martens, P.~C.~H.\ 1989, \apj, 343, 971
\bibitem[Xia et al.(2011)]{xia11} Xia, C., Chen, P.~F., Keppens, R., \& van
	Marle, A.~J.\ 2011, \apj, 737, 27 
\bibitem[Xia et al.(2012)]{xia12} Xia, C., Chen, P.~F., \& Keppens, R.\ 2012,
	\apjl, 748, L26 
\bibitem[Yan et al.(2008)]{yan08} Yan, X.~L., Qu, Z.~Q., \& Xu, C.~L.\ 2008,
	\apjl, 682, L65
\bibitem[Zhang \& Low(2005)]{zhan05} Zhang, M., \& Low, B.~C.\ 2005, \araa,
	43, 103
\bibitem[Zhang, Fang, Zhang(2012)]{zhang12} Zhang, P., Fang, C., \& Zhang, Q.\
	2012, Science in China G: Physics and Astronomy, 55, 907
\bibitem[Zhang et al.(2012)]{zhan12} Zhang, Q.~M., Chen, P.~F., Xia, C., \&
	Keppens, R.\ 2012, \aap, 542, A52 
\bibitem[Zhang et al.(2013)]{zhan13} Zhang, Q.~M., Chen, P.~F., Xia, C.,
	Keppens, R., \& Ji, H.~S.\ 2013, \aap, 554, A124 
\bibitem[Zirker(1989)]{zirk89} Zirker, J.~B.\ 1989, \solphys, 119, 341
\end{thebibliography}
\end{document}